\documentstyle[12pt,epsfig]{article}
\def\bfk{\mbox{\boldmath $k$}}

\def\be{\begin{equation}}
\def\ee{\end{equation}}
\def\bea{\begin{eqnarray}}
\def\eea{\end{eqnarray}}
\def\nd{\noindent}
\begin{document}

\begin{center}
{\bf What do we know about the proton spin structure? \footnote{Talk 
delivered at the 2nd International Symposium on the Gerasimov-Drell-Hearn
sum rule and the spin structure of the nucleon, GDH 2002, July 3-6 2002,
Genova, Italy}}

\vskip 0.8cm
{\sf Mauro Anselmino}
\vskip 0.5cm
{\it Dipartimento di Fisica Teorica, Universit\`a di Torino and \\
    INFN, Sezione di Torino, Via P. Giuria 1, I-10125 Torino, Italy}\\
\end{center}

\vspace{1.5cm}

\begin{abstract}
A brief summary of the theoretical and experimental knowledge 
of the spin structure of the proton is presented. The helicity distributions 
of quark and gluons are discussed, together with their related sum rules.  
The transversity distribution is also introduced with possible strategies 
for its measurement. Novel spin dependent and $\bfk_\perp$ unintegrated 
distribution and fragmentation functions are discussed, in connection with a 
new and rich phenomenology of transverse single spin asymmetries.
\end{abstract}

\nd 
{\bf Introduction} 
\vskip 6pt 

The spin nucleon structure -- as observed in high energy, short distance
interactions -- is schematically described in Fig. 1.
\begin{figure}[ht]
\centerline{\epsfxsize=3.9in\epsfbox{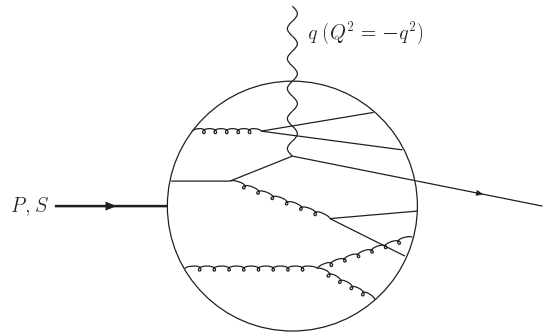}}   
\caption{The nucleon structure, as seen by a high energy, large $Q^2$, probe.
The nucleon breaks into unobserved final particles $X$; the ejected parton 
may fragment into an observed hadron.}
\label{fig1}
\end{figure}

The large $Q^2$ probe -- typically, a virtual photon -- ``sees'' QCD partons,
carrying a longitudinal momentum fraction $x$, and their interactions, with 
gluon and $q\bar q$ pair creation; the information about such a complicated 
structure is usually collected via measurements of the Deep Inelastic 
Scattering (DIS) cross-section and storaged in the structure functions 
which appear in the most general expression of the cross-section. 
When neglecting weak, parity violating contributions, there 
are 2 unpolarized ($F_1$, $F_2$) and 2 polarized ($g_1$, $g_2$) structure 
functions: perturbative QCD allows a simple partonic interpretation of $F_2$
and $g_1$ ($F_1$ is related to $F_2$ while $g_2$ does not have a partonic
interpretation). We only consider here the polarized proton structure trying
to summarize in a short time and space the main ideas, the most recent 
results and the open problems; many detailed and comprehensive reviews on 
the subject can be found in the literature \cite{rev}.       

The main issues and questions we are going to discuss here are:

\begin{itemize}
\item
our knowledge about the polarized structure functions $g_1(x,$ $Q^2)$ 
and $g_2(x,Q^2)$ and about quark and gluon helicity distributions, 
$\Delta q(x,Q^2)$ and $\Delta g(x,Q^2)$; how well do we know them?
\item
are fundamental sum rules satisfied and what do we know about quark and 
gluon orbital angular momentum, $L_q$ and $L_g$?
\item
$\Delta q(x,Q^2)$, $\Delta g(x,Q^2)$, $L_q$ and $L_g$ are not the whole 
story: how and where do we learn about the transversity distribution 
$h_1(x,Q^2)$?
\item
could we learn more and understand more from intrinsic $\bfk_\perp$
{\it unintegrated} distribution and fragmentation functions?  
\end{itemize}

\nd 
{\bf The longitudinally polarized proton and the helicity distributions}
\vskip 6pt 

At NLO in the QCD parton model the structure function $g_1$ is given by
\be
g_1(x, Q^2) = 
\frac 12 \sum_q e_q^2 \left\{ \Delta C_q \otimes \left[ \Delta q + 
\Delta\bar q \right] + \frac {1}{N_f} \Delta C_g \otimes \Delta g \right\} 
\label{g1evol}
\ee
where $\Delta q(x, Q^2)$ and $\Delta g(x, Q^2)$ are respectively 
the quark (of flavour $q$) and gluon helicity distributions; we 
have, as usual, defined the convolution
\be 
\Delta C \otimes \Delta q \equiv \int_x^1 \frac {dy}{y} \> 
\Delta C \!\! \left( \frac xy, \alpha_s \right) \> \Delta q(y, Q^2)
\label{conv}
\ee
and the coefficients functions $\Delta C_i$ have a perturbative
expansion
\be
\Delta C_i(x, \alpha_s) = \Delta C_i^0(x) + \frac {\alpha_s(Q^2)}{2\pi} \,
\Delta C_i^{(1)}(x) + \cdots \label{coeff} 
\ee

The LO terms are simply
\be
\Delta C_q^0 = \delta (1-x) \quad\quad\quad\quad \Delta C_g^0 = 0 \,,
\label{lo}
\ee
and the NLO corrections are scheme dependent; typical choices differ in 
the amount of gluon contribution to the quark singlet distributions, while 
quark non-singlet distributions are scheme independent \cite{rev}. Finally, 
the $Q^2$ evolution of the parton densities obeys the DGLAP evolution 
equations \cite{dglap}, and, if known at an initial scale $\mu^2$, the 
r.h.s. of Eq. (\ref{g1evol}) can be computed at any perturbative $Q^2$ value. 

By comparing data on $g_1(x, Q^2)$ with Eq. (\ref{g1evol}) one obtains 
information on the quark and gluon helicity distributions; the more data 
one has and the wider the $x$ and $Q^2$ range is, the more stringent the 
comparison is. The normal procedure is that of using a simple ansatz for 
the unknown distribution functions at the initial scale $\mu^2$, with some 
assumptions regarding the sea quark densities (for example, whether 
$SU(3)_F$ symmetric or not) and some constraints from $SU(3)_F$ hyperon 
decay sum rules on the first moments 
$\Delta q(1,Q^2) \equiv \int_0^1 \Delta q(x,Q^2) \> dx$.

In Fig. 2 a most recent analysis of the world data on $xg_1(x)$ is shown 
together with a fit from Ref. [3], where the resulting helicity 
distributions can also be seen. Several similar analyses can be found in 
the literature; a complete list of references is given in Ref. [3]. 
\begin{figure}[ht]
\centerline{\epsfxsize=3.5in\epsfbox{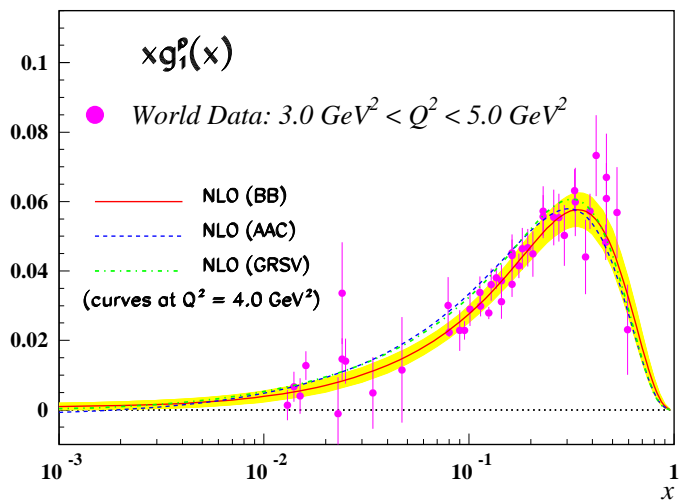}}   
\caption{The world data on $xg_1^p(x)$; the figure is taken from Ref. [3]}
\label{fig2}
\end{figure}
 
Some data on $xg_2(x)$ are also available and the most recent ones \cite{e155} 
are shown in Fig. 3. 

\begin{figure}[ht]
\centerline{\epsfxsize=3.5in\epsfbox{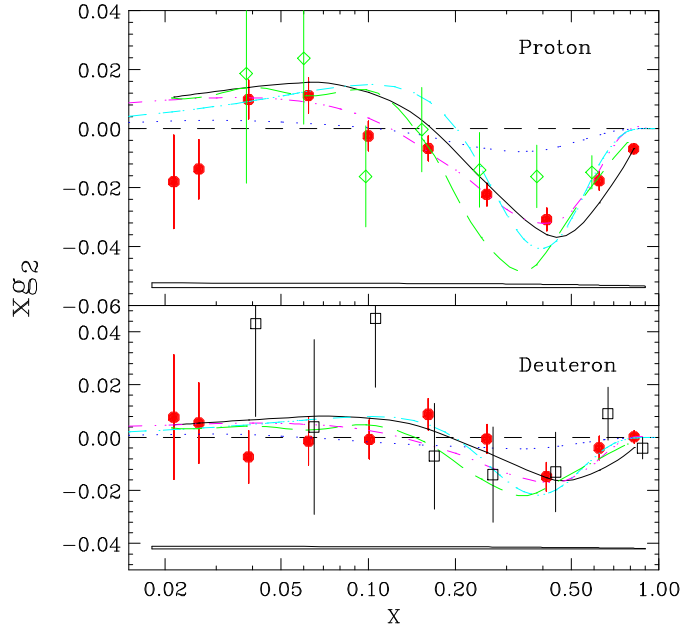}}   
\caption{The E143 and E155 data on $xg_2^p(x)$ and $xg_2^d(x)$, 
from Ref. [4].}
\label{fig3}
\end{figure}

Let us shortly comment on these experimental results and the information 
which they offer.

\begin{itemize}
\item
We have now good data on $g_1$ and $g_2$, although not yet comparable with
the amount and quality of similar data obtained on the unpolarized structure 
functions.
\item
$g_1$ [see Eq. (\ref{g1evol})] allows to obtain information on linear 
combinations of ($\Delta q + \Delta \bar q$). We still need a better flavour 
resolution; this might come from semi-inclusive DIS which gives information
on $\sum_q \, \Delta q \, D_q^h$ rather than $\sum_q e_q^2 
(\Delta q + \Delta \bar q)$, where $D_q^h$ is the quark $q$ fragmentation 
function into the observed hadron $h$. Flavour separation in inclusive DIS 
would naturally be possible in neutrino iniziated charged current 
processes \cite{fmr}.
\item
Eq. (\ref{g1evol}) also offers indirect (via QCD evolution) information, 
on $\Delta g$. This is not stringent enough and a more direct 
measurement of the gluon helicity distribution is needed. This might come
from the study of spin dependences in processes like $\ell p \to \ell + 
2\,jets$, $\ell p \to \ell + c + \bar c + X$, $p N \to \gamma + X$, {\it etc.} 
which could be performed at HERMES, COMPASS, RHIC.  
\end{itemize}
 
\nd
{\bf Sum rules and orbital angular momentum}
\vskip 6pt

In extracting information from experimental data -- or in testing theories -- 
a special role is plaid by sum rules. Let us mention a few of them.

The Bjorken sum rule ($g_A/g_V = 1.2670 \pm 0.0035$),
\bea
&& \int_0^1 [g_1^p(x,Q^2) - g_1^n(x,Q^2)]\,dx  \label{bj} \\
&=& \frac 16 \frac{g_A}{g_V} \left\{ 1 - \frac{\alpha_s}{\pi} - \frac {43}{12}
\,\frac{\alpha_s^2}{\pi^2} - 20.2\,\frac{\alpha_s^3}{\pi^3} + 
\cdots \right\} \>,
\nonumber   
\eea
is used in many ways. One can simply assume the validity of Eq. (\ref{bj})
and deduct from data on $g_1$ the value of $\alpha_s$ \cite{abfr}; or one 
can use in it a value of $\alpha_s$ otherwise obtained, to check whether 
data on $g_1$ obey the sum rule or not (the answer is yes); 
or, also \cite{smallx}, one can assume to know the r.h.s. of Eq. (\ref{bj}),
and see -- among the poorly known behaviours of $g_1(x)$ at small $x$ -- 
which one best satisfies the sum rule.
    
Another, more debated, sum rule is the so called Burkhardt-Cottingham sum 
rule, according to which $\int_0^1 g_2(x,Q^2)\,dx = 0$, {\it provided} the 
integral exists. The recent E155 data \cite{e155} of Fig. 3 seem to indicate 
$\int_{0.02}^{0.8} g_2^p(x,Q^2)\,dx = - 0.042 \pm 0.008$, which, taking into 
account uncertainties in the extrapolation to $x=0$ and $x=1$, might be the 
first indication of a violation of the sum rule, assuming that no 
$\delta$-function contributes at the origin \cite{del}.  
 
The last, fundamental sum rule which we mention is the spin sum rule:
\be 
\frac 12 = 
\frac 12 \Delta \Sigma(1) + \Delta g(1) + L_q + L_g \label{spin}
\ee
where $\Delta \Sigma(1)$ is the first moment of $\sum_q 
[\Delta q(x) + \Delta \bar q(x)]$ and $L_{q,g}$ is the third component 
of the orbital angular momentum carried by quarks, gluons.
This last quantity is unavoidable in a picture of the proton like 
that of Fig.~1: a spin 1/2 massless quark can emit a spin 1 massless gluon, 
via a helicity conserving coupling, only if some orbital angular momentum 
restores the total angular momentum conservation.
However, there is little agreement at the moment both about the proper formal 
definition of a $\hat L_{q,g}$ operator and about a possible 
measurement of its expectation value between proton states \cite{jaf}. 

\vskip 8pt
\nd       
{\bf Transversity distribution}
\vskip 6pt

The trasverse polarization of quarks inside a trasversely polarized nucleon,
denoted by $h_1$, $\delta q$ or $\Delta_T q$, is a fundamental twist-2 
quantity, as important as the unpolarized distributions $q$ and the 
helicity distributions $\Delta q$. It is given by
\be 
h_1(x, Q^2) = q_\uparrow^\uparrow(x, Q^2) - q_\downarrow^\uparrow(x,Q^2) \>,
\ee
that is the difference between the number density of quarks with transverse
spin parallel and antiparallel to the nucleon spin. It is the same as the
helicity distribution only in a non relativistic approximation,
but it is expected to differ from it for a relativistic nucleon.  

When represented in the helicity basis (see Fig. 4) $h_1$ relates quarks 
with different helicities, revealing its chiral-odd nature. This is the 
reason why this important quantity has never been measured in DIS:
the electromagnetic or QCD interactions are helicity conserving, there
is no perturbative way of flipping helicities and $h_1$ decouples from
inclusive DIS dynamics, as shown in Fig. 4a.

However, it can be accessed in semi-inclusive DIS, where some
non perturbative chiral-odd effects may take place in the non perturbative 
fragmentation process, Fig. 4b. Indeed, a serious program to measure $h_1$ 
in semi-inclusive DIS at HERMES, where a transversely polarized proton
target is now available, is in progress. A similar program, in different,
complementary, kinematical regions, is planned at COMPASS.  
\begin{figure}[ht]
\centerline{\epsfxsize=3.5in\epsfbox{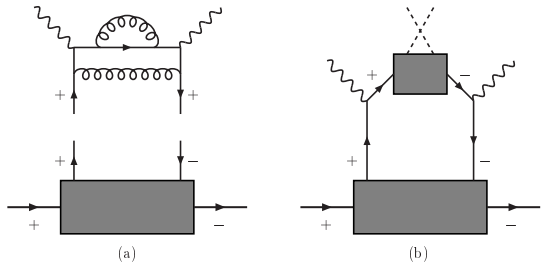}}   
\caption{The chiral-odd function $h_1$ (lower box) cannot couple to inclusive
DIS dynamics, even with QCD corrections; it couples to semi-inclusive DIS,
where chiral-odd non perturbative fragmentation functions may appear.}
\label{fig4}
\end{figure}

The transversity distribution is also accessible at RHIC, where 
transversely polarized proton beams are available; by measuring double
transverse spin asymmetries in Drell-Yan processes one obtains an
observable which depends on the convolution of two transversity 
distributions, which might make the overall effect rather tiny \cite{rhic}. 
In general, $h_1$ must appear in a physical observable coupled to another
chiral-odd quantity, which is either the transversity itself or a new
unknown function. 

\vskip 8pt
\vskip 8pt
\nd
{\bf Unintegrated distribution and fragmentation functions}
\vskip 6pt

We conclude by mentioning a new phenomenological approach to the description 
of many single transverse spin asymmetries which have been measured and keep
being measured, with unexpected and interesting results \cite{me}.
The apparent problem with these asymmetries is related to the fact that, 
within perturbative QCD and the collinear factorization scheme, they should 
be vanishing, which is not true experimentally. 

Recently, a series of papers \cite{asy} have shown how single spin 
asymmetries may occurr at the level of parton distributions and 
fragmentations, provided one takes into account the intrinsic motion of 
partons inside hadrons and of hadrons relatively to the fragmenting parton. 
For example, there might be a correlation between the transverse spin of 
a quark and the $\bfk_\perp$ of a resulting hadron, say a pion. This is the 
so-called Collins effect \cite{col}, pictorially shown in Fig. 5. 
\begin{figure}[ht]
\centerline{\epsfxsize=3.5in\epsfbox{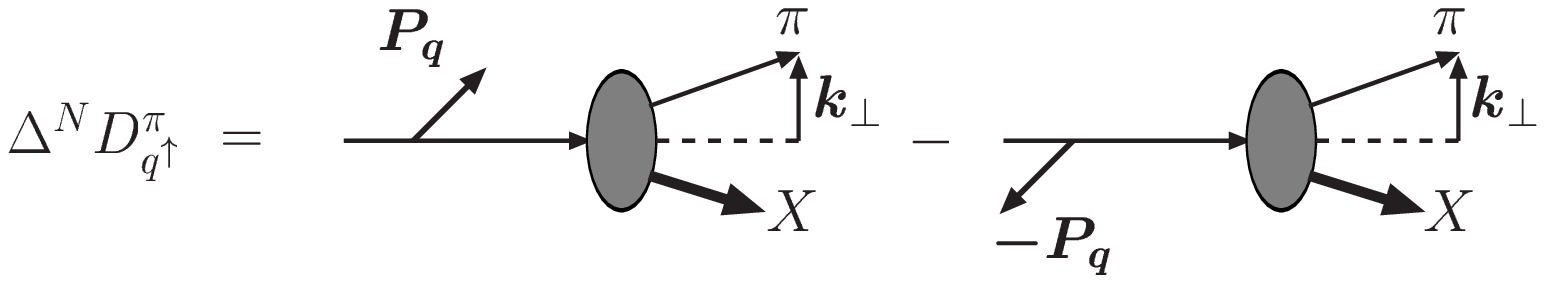}}   
\caption{Pictorial representation of Collins function; notice that a similar
function is sometimes denoted by $H_1^\perp$ in the literature.}
\label{fig5}
\end{figure}

Similar spin-$\bfk_\perp$ correlations may occurr also in the fragmentation 
of an unpolarized quark into a polarized hadron (the so-called polarizing 
fragmentation functions \cite{polff}), in the distribution of unpolarized 
quarks inside polarized nucleons (the Sivers effect \cite{siv}) and in the 
distribution of polarized quarks inside an unpolarized hadron \cite{dan}. 

When generalizing the factorization scheme with the inclusion of intrinsic
$\bfk_\perp$, both in the distribution/fragmentation functions and in the 
elementary interactions, single transverse spin asymmetries appear 
immediately as possible and even sizeable. A phenomenological approach 
can be developed in which experimental information on the new functions is 
obtained from some processes and then used to make predictions in other 
cases.       

The spin structure of the nucleon is subtle and challenging.
Enormous progress has been achieved in the last years; yet, new surprising
experimental results keep beeing obtained and fresh, interesting ideas keep
being suggested. A lot more good work, both experimental and theoretical,
is in progress.   
 
\vskip 12 pt
\nd   
{\bf Acknowledgments}
I would like to thank the organizers of the Symposium for the invitation 
and for the beautiful and stimulating organization.

\end{document}